\documentclass[10pt,twocolumn]{article}
\pdfoutput=1

\usepackage{Dincon}

\title{\uppercase{Dynamical aspects of Kinouchi-Copelli model:\\
emergence of avalanches at criticality}}

\author[1,$\dagger$]{\underline{T. S. Mosqueiro}}
\author[1,$\S$]{C. Akimushkin}
\author[1,$\ddagger$]{L. P. Maia}

\affil{Instituto de F\'{\i}sica de S\~ao Carlos (USP), S\~ao Carlos, Brasil}

\affil[$\dagger$]{e-mail: thiago.mosqueiro@gmail.com}
\affil[$\S$]{e-mail: camilo.akimushkin@gmail.com}
\affil[$\ddagger$]{e-mail: lpmaia@ifsc.usp.br}
\affil[ ]{\bf Paper accepted for oral presentation}

\begin{document}

\maketitle

\Abstract{We analyze the behavior of bursts of neural activity in the Kinouchi-Copelli model, originally conceived to explain information processing issues in sensory systems. We show that, at a critical condition, power-law behavior emerges for the size and duration of the bursts (avalanches), with exponents experimentally observed in real biological systems.}

\Keywords{Neuro-dynamics, critical phenomena, avalanches}

\section{INTRODUCTION}

The theory of critical phenomena in equilibrium statistical physics became one of the landmarks of Physics in the last century (culminating with the Nobel prize for Kenneth Wilson in 1982) because, among other things, it offered an explanation for the phenomenon of universality, where physical systems in principle very distinct, like a magnet and a fluid, can present the same behavior (even in a quantitative sense) in some key quantities when they are all close to a so-called critical point. Namely, the magnetization in some model magnets and the difference of density of gas and liquid phases in a fluid both behave as power-laws \textit{with the same exponent} as functions of the temperature when this control parameter is close enough to a critical value.

The ubiquity of power-laws in nature suggested that the concept of criticality could be extended to nonequilibrium settings and then become relevant also to the understanding of phenomena in domains outside ``regular'' physics, like biological and geophysical systems, for instance. Nevertheless there is still no general theoretical formalism for such ``nonequilibrium-criticality'' and there is few (if any) uncontroversial experimental evidence of a role of criticality in the nonequilibrium processes observed in nature. One research avenue where such evidence could rise is the study of brain dynamics.

In the last decade, experiments with cortical networks both \textit{in vivo} and \textit{in vitro} \cite{Beggs03122003} have revealed activation patterns of neurons characterized by power-law distributions in the number of units that got eventually excited as well as in the total duration of a burst of activity. We will call such processes as neuronal avalanches or simply \textit{avalanches}, in spite of the fact that other researchers use that expressions for any pattern of neural activity (what obliges them to introduce additional qualifiers as ``scale-free'', ``critical'' or ``power-law distributed'' when they talk about the avalanches we are interested in this paper).

Those experiments were interpreted as manifestations of self-organized criticality (SOC) \cite{PhysRevLett.59.381}, a theoretical proposal for general nonequilibrium critical behavior by which the interactions among the units of complex systems would ``naturally'' evolve in time towards a condition where the asymptotic spatial-temporal activity would be characterized by power-law distributions. It is important to note the absence of a human-tuned control parameter as temperature in the equilibrium critical phenomena. It would be hard to conceive an alternative framework that looks more suited to describe the synaptic dynamics of neural systems. Even so, despite appealing, SOC received many criticisms because of the lack of any precise description of the general mechanisms of the underlying adaptive process and also because of the lack of an unambiguous signature, both in neuroscience and other applications. Indeed, many recent works have focused in exhibiting scale-free behavior without any explicit adaptive SOC-like mechanism \cite{benayounploscompbio7e1000486}.

In this paper we analyze the avalanches that arise in a model \cite{kinouchicopellinaturephys23482006} conceived by Osame Kinouchi and Mauro Copelli (KC hereafter) to explain how sensory systems could be able to discriminate signals spanning many orders of magnitude in intensity. This interval of values of a signal discriminable by a system is called dynamic range. Remarkably, KC found out that at least some classes of networks of excitable elements may exhibit an optimal (larger) dynamic range at a critical condition. However, it should be stressed that such a nonequilibrium phase transition is not self-organized at all (there is no adaptive process and a proper control parameter must be ``tuned''). Experiments with cortical slices even inspired speculations regarding a possible causal connection between avalanches and an optimal dynamic range \cite{shewjon292009}. Anyway, the authors of \cite{kinouchicopellinaturephys23482006} have made some comments on the nature of the avalanches in their model but, to the best of our knowledge, have not publicly characterized that critical bursts, what we do in this study.

This paper is organized as follows. We define the Kinouchi-Copelli model in section \ref{sec:kc}. We exhibit the avalanches (size and time) probability distributions in section \ref{sec:avalanches}.  In section \ref{sec:conc} we summarize our results and indicate directions for further investigations.

\section{KINOUCHI-COPELLI MODEL}
\label{sec:kc}

Consider an undirected weighted Erd\"{o}s-R\'{e}nyi random graph with $N$ nodes. Each node represent a neuron, \textit{i.e.} an excitable unit whose possible states will be described below. Given a desired average connectivity $K$ (mean degree of a node) for the graph, each of the $NK/2$ edges is assigned to a randomly chosen pair of nodes. Let $V_j$ be neighborhood of node $j$, \textit{i.e.} the set of nodes connected to $j$ by an edge. The strength of a synapse in this neuronal network is represented by the weight of the corresponding edge in this graph and is sorted from an uniform probability density in the interval $[0,p_{max}]$, where $0 \leq p_{max} \leq 1$. Representing the absence of a synapse as a null weight, we can define an (symmetric) adjacency matrix $A$ whose element $A_{jk}$ is the weight of the edge between nodes $j$ and $k$.

Let $X_j (t)$ be a random variable representing the state of the $j$-th neuron at the instant $t$. For all $j$ and $t$, $X_j (t)$ exhibits realizations in the set $\{0,1,\cdots,m-1\}$. The state $0$ is called either the \textit{quiescent state} or the rest state. The state 1 is the \textit{excited state} and all other states are called \textit{refractory states}. The full dynamics of the system consists in the temporal evolution of the family $\{ X_j (t) \}$ in discrete time, with synchronous updating, according to the following rules:
\begin{itemize}
  \item if $1 \leq X_j(t) \leq m-2$, then $X_j(t+1) = X_j(t) + 1$;
  \item if $X_j(t) = m-1$, then $X_j(t+1) = 0$;
  \item if $X_j(t) = 0$, then $X_j(t+1)$ will be either 0 or 1, and the total excitation probability of neuron $j$ depends on independent contributions from
  \begin{itemize}
    \item an external stimulus with probability $\eta$;
    \item each of its excited neighbors, say $k$, with probability $A_{jk}$.
  \end{itemize}
\end{itemize}
Explicitly, ``independent contributions'' mean that each of the numbers $\eta$ and $\{A_{jk}\}$ are meaningful as excitation probabilities only in isolation (absence of all other contributions). Also notice that the refractory period of a neuron equals $m-1$ time steps, starting right after this neuron getting excited. Its evolution is deterministic meanwhile. The only probabilistic state transition occurs from the quiescent state to the excited one.

For KC $\eta = 1-e^{-r \Delta t}$, where $\Delta t$ would be an arbitrary continuous time interval (usually, $\Delta t \approx 1 \ ms$) and $r$ would be the probability rate of a Poisson process. In olfactory intraglomerular neuronal networks (a biological system where the KC model may be applicable), $r$ would be directly related to the concentration of an odorant capable of exciting neurons.

\begin{figure}
  \includegraphics[scale = 0.18]{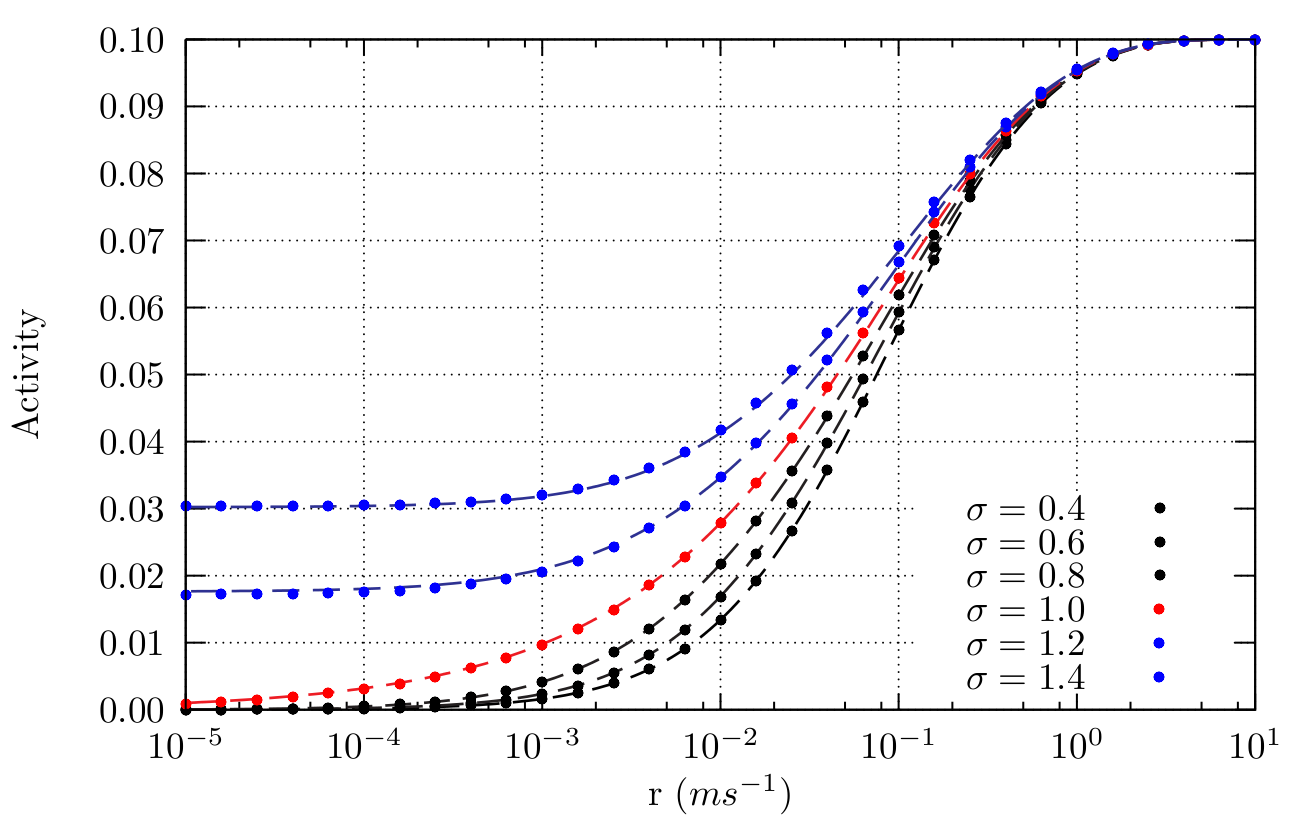}
  \caption{Typical behaviour of the activity $F$ as a function of the excitation rate $r$. The dashed line is a mean-field solution proposed by KC \cite{kinouchicopellinaturephys23482006}.}
\label{fig:F_x_r}
\end{figure}

The main observables of the KC model are the density of excited neurons $\rho(t)$ at $t$-th time step (the fraction of the population of neurons composed by excited units) and its temporal average, the activity of the network,
\begin{equation}
  F := \frac{1}{T} \sum\limits_{t = 1}^T \rho(t).
  \label{eq:activity}
\end{equation}
For large enough ($\approx 10^3 \ ms$) values of the observation window $T$, so that a dynamical equilibrium is reached, its precise value does not have relevant effects on the behavior of $F$. Then the activity can be seen as a function of the excitation rate $r$ as shown in Fig. \ref{fig:F_x_r}.

The critical behavior is revealed only when $\Delta r ^*$, the range of values of $r$ over which $F$ exhibits ``significant'' variation \cite{kinouchicopellinaturephys23482006}, is seen as a function of the average branching ratio $\sigma$, defined as the mean value (averaged over all the neurons) of the local branching ratio $\sigma_j$ of the $j$-th node,
\begin{equation}
\sigma_j
=
\sum_{k \in V_j} A_{jk}.
\label{eq:sigmaj}
\end{equation}
Indeed, the dynamic range $\Delta r ^*$ turns out to be optimal when $\sigma = 1$. So the role of control parameter is performed by $\sigma$, which is a measure of how much activity can be directly generated by an excited unit of the network stimulating a resting neighborhood.

\section{\uppercase{Emergence of avalanches at criticality}}
\label{sec:avalanches}

It is not our interest here to discuss the optimal dynamic range, despite that being the major result in \cite{kinouchicopellinaturephys23482006}. Instead, we are concerned with the dynamical properties of the KC model when there is a non-null initial density of excited neurons and the external stimulus is turned off. As already discussed in \cite{kinouchicopellinaturephys23482006}, for  $\sigma \leq 1$, $F = 0$, \textit{i.e.}, any given signal (initial condition) is damped; for $\sigma > 1$, $F \neq 0$, \textit{i.e.}, the signal is self-sustained by the network. How the patterns of neural activity depend on $\sigma$? KC already pointed out there are avalanches when $\sigma=1$ but did not characterize such bursts. In the following, we discuss the methodology we have employed to characterize the avalanches and present the distributions of size and duration of that events.

\subsection{Methodology}

To probe the capability of the KC network in propagating a signal, we induce a burst of activity by exciting a randomly chosen single neuron and keeping the external stimulus turned off. The duration of a burst is the number of time steps in which there is at least one active unit. The burst size is the total number of neurons that got eventually excited (avoiding multiple counts). We repeat this procedure a great number of times (typically $2 \cdot 10^5$) to generate the data we analyze in the following subsections.

There, we exhibit the probability density functions (PDF) and the cumulative distribution functions (CDF) for both size and duration of bursts. Clearly, the PDF's in the critical condition of the KC model ($\sigma=1$) are power laws. It is easy to see that a CDF of such a PDF is a power law too. We have used the CDF's to estimate the power-law exponents as is usually recommended \cite{newmancp2005,clausetsiamr2009}. We have not shown any supercritical curves because they are as different from the critical behavior as are the subcritical data and therefore they would not be illustrative at all.

\subsection{Distribution of avalanche sizes}

\begin{figure}[h]
  \includegraphics[width=8cm,height=5cm]{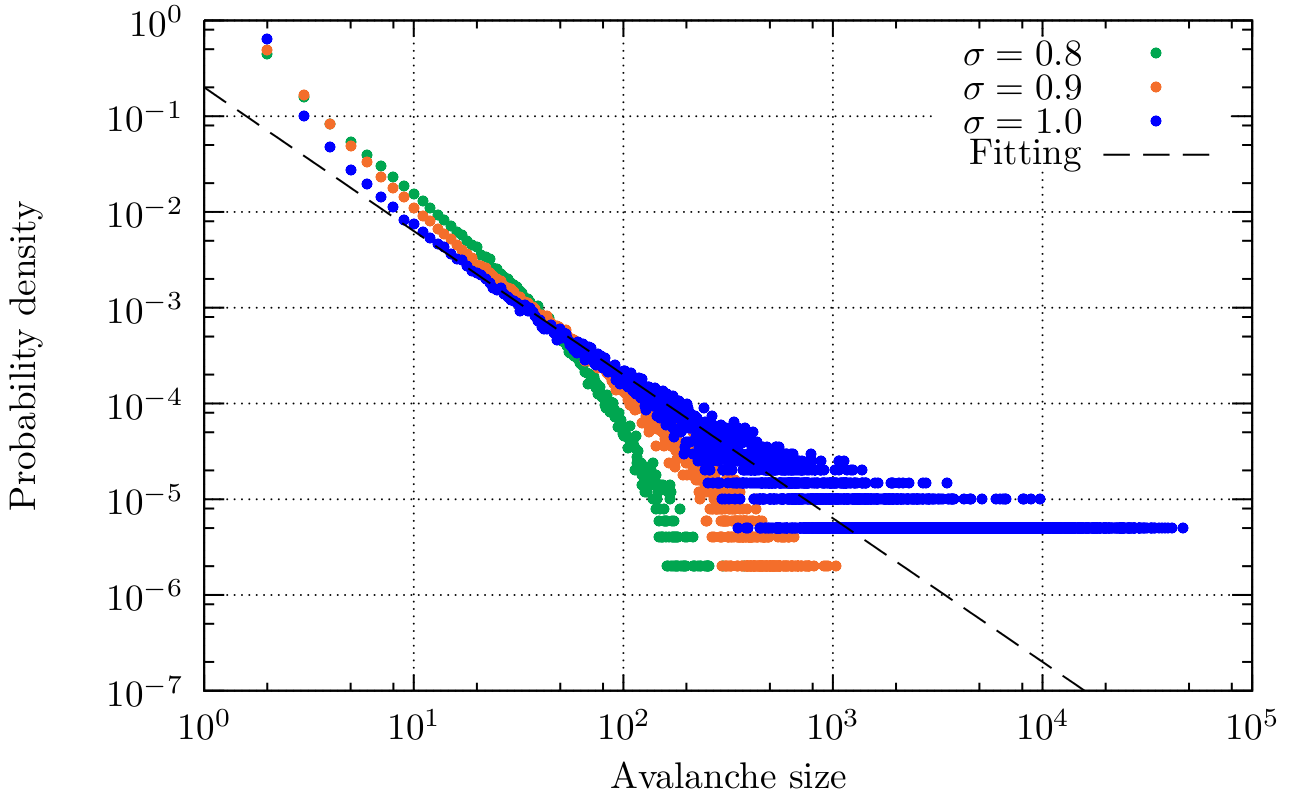}
  \caption{PDF of avalanche size for three different values of $\sigma$. The power law emerges only in the critical condition ($\sigma=1$). Data from networks with $10^5$ neurons, $m = 10$ refractory states and mean connectivity $K=10$.}
    \label{fig:size_sub}
\end{figure}

\begin{figure}[h]
  \includegraphics[width=8cm,height=5cm]{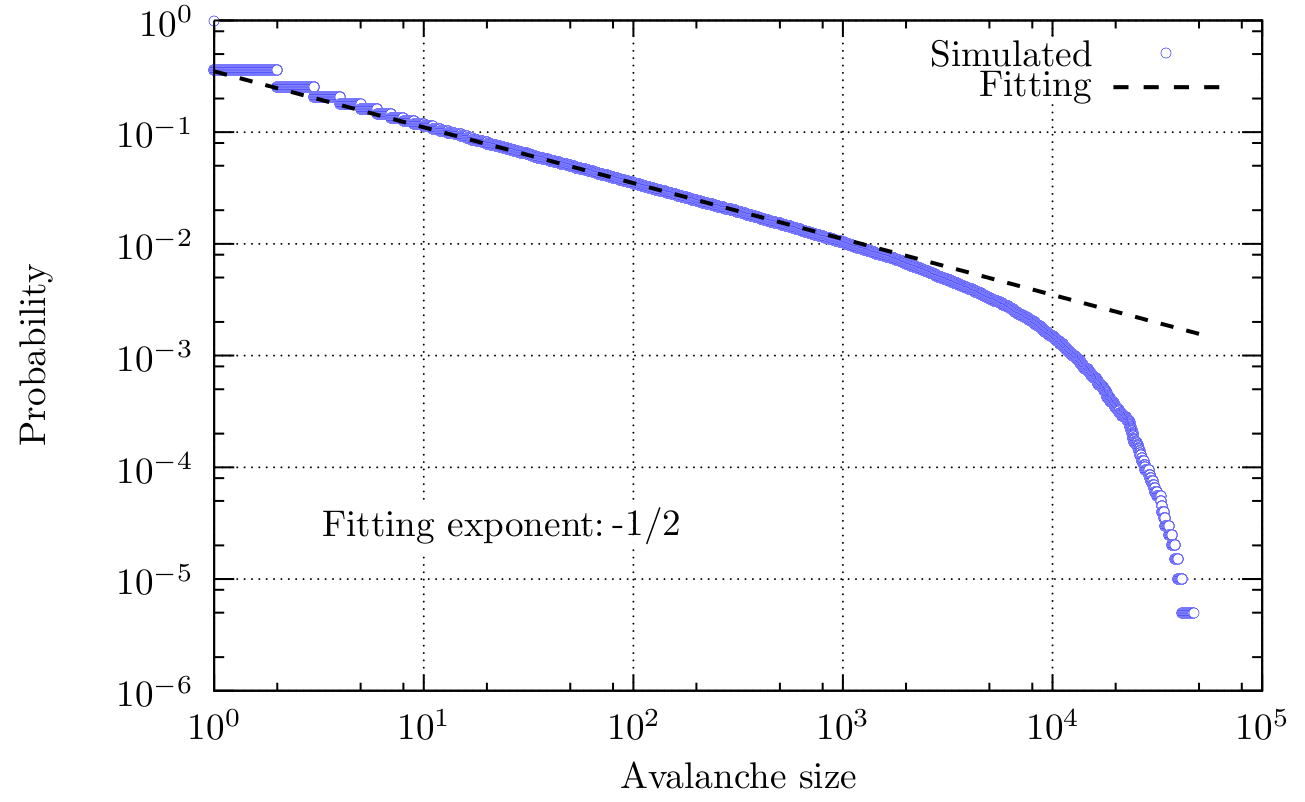}
  \caption{CDF of avalanche size in the critical condition ($\sigma=1$). The power law behavior extends for almost four orders of magnitude. Data from networks with $10^5$ neurons, $m = 10$ refractory states and mean connectivity $K=10$.}
    \label{fig:size_cdf}
\end{figure}

The PDF's for three values of the average branching ratio $\sigma$ are shown in Fig. \ref{fig:size_sub}. It is reasonable to say that we have a power law only in the critical case, but it becomes ``crystal clear'' in Fig. \ref{fig:size_cdf}. There we see an exponent close to $-1/2$ that indirectly implies an exponent close to $-3/2$ for the PDF. This corroborates a comment by KC \cite{kinouchicopellinaturephys23482006} and agrees with experimental measurements \cite{shewjon292009}. It is important to notice that the power-law behavior in Fig. \ref{fig:size_cdf} extends for almost four orders of magnitude, a very robust signature indeed.

\subsection{Distribution of avalanche times}

\begin{figure}[h]
  \includegraphics[width=8cm,height=5cm]{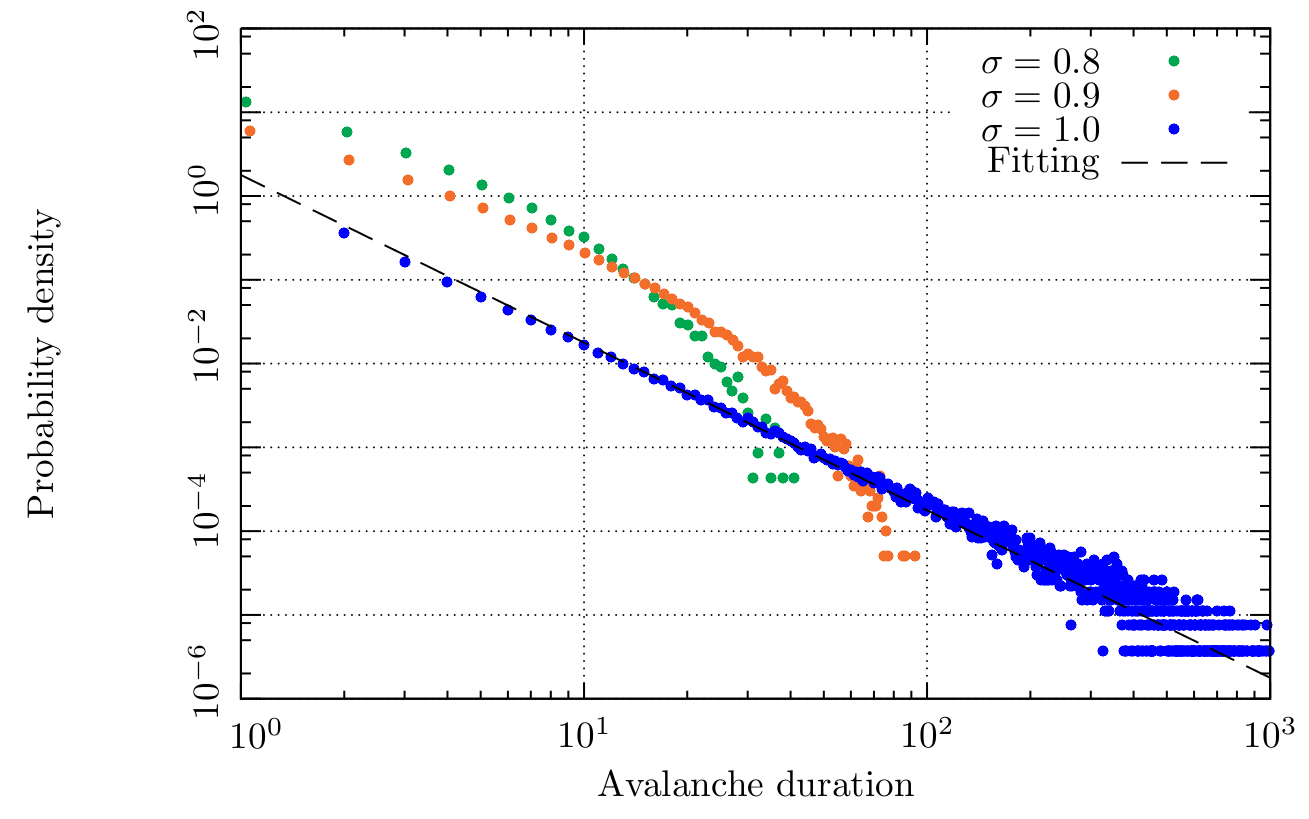}
  \caption{PDF of avalanche time for three different values of $\sigma$. The power law emerges only in the critical condition ($\sigma=1$). Data from networks with $10^5$ neurons, $m = 10$ refractory states and mean connectivity $K=10$.}
    \label{fig:time_sub}
\end{figure}

\begin{figure}[h]
  \includegraphics[width=8cm,height=5cm]{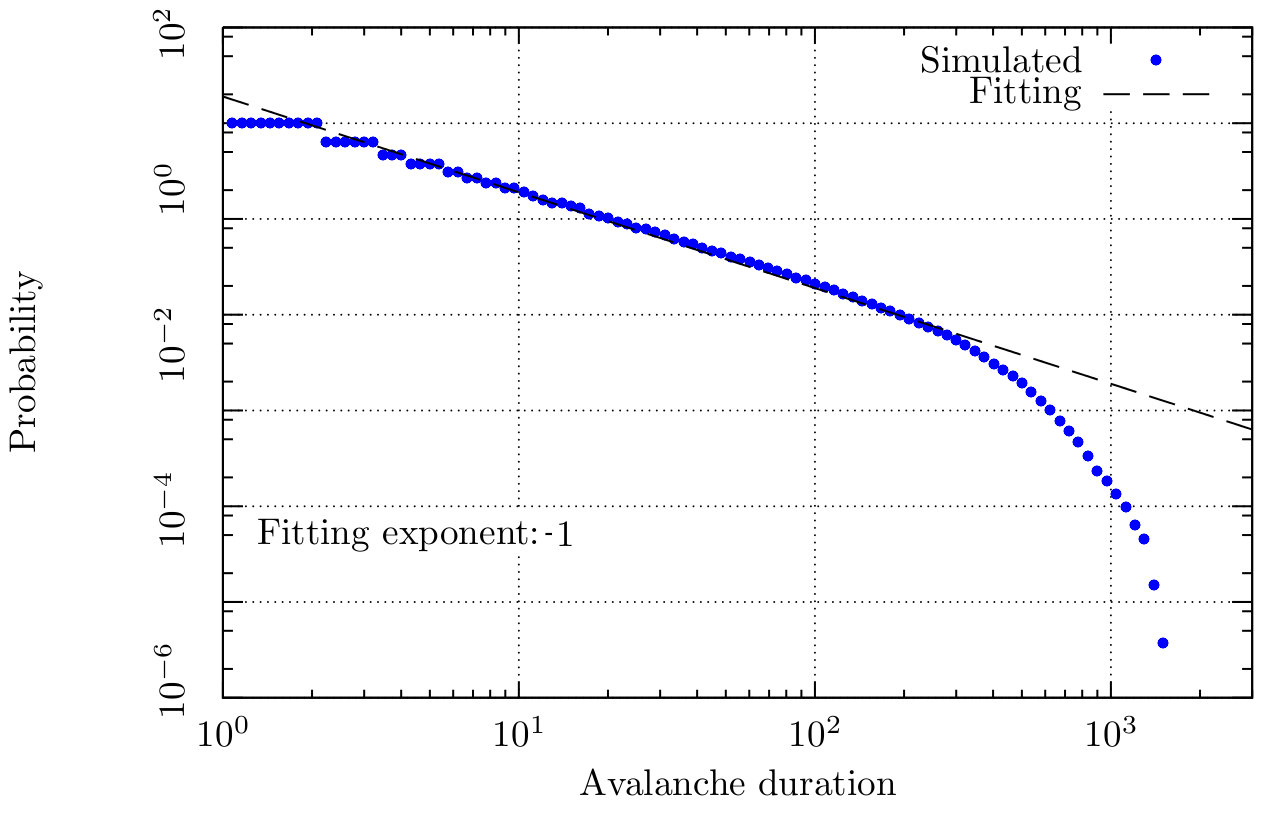}
  \caption{CDF of avalanche time in the critical condition ($\sigma=1$). The power law behavior extends for almost four orders of magnitude. Data from networks with $10^5$ neurons, $m = 10$ refractory states and mean connectivity $K=10$.}
    \label{fig:time_cdf}
\end{figure}

The analysis is performed analogously to the case of size distribution. The results are shown in Figs. \ref{fig:time_sub} and \ref{fig:time_cdf}. We extract an exponent close to $-2$ for the PDF from the exponent $-1$ in the corresponding CDF. This is in agreement with the experimental result of J. Beggs and D. Plenz \cite{Beggs03122003,Beggs02062004} that initiated the search for criticality in brain dynamics and indirectly motivated the present paper. That fact was not reported in \cite{kinouchicopellinaturephys23482006}. Finally we notice that the power-law behavior in Fig. \ref{fig:time_cdf} extends only for almost three orders of magnitude.

\section{CONCLUSION AND FUTURE DIRECTIONS}
\label{sec:conc}

We have analyzed the avalanche behavior in the Kinouchi-Copelli model. It reveals the exponents $-3/2$ and $-2$ for the power laws characterizing the size and duration of the avalanches, respectively. It is noteworthy that both exponents have already been experimentally observed \cite{shewjon292009,Beggs03122003,Beggs02062004,10.1371/journal.pone.0000439}. Therefore our study fits in the search for criticality in brain dynamics, despite not being related to self-organized criticality. In the near future we intend to investigate: (i) the effect of different network topologies, (ii) finite-size effects and (iii) how the properties of avalanches would be affected by the competition among avalanches that certainly occurs in real biological systems.

\section*{ACKNOWLEDGMENTS}

We acknowledge CAPES and FAPESP for financial support.


\bibliographystyle{unsrt}
\bibliography{/home/thmosqueiro/Dropbox/bibtex/articlesphys,/home/thmosqueiro/Dropbox/bibtex/booksphys}

\begin{thebibliography}{1}

\bibitem{Beggs03122003}
John~M. Beggs and Dietmar Plenz.
\newblock Neuronal avalanches in neocortical circuits.
\newblock {\em The Journal of Neuroscience}, 23(35):11167--11177, 2003.

\bibitem{PhysRevLett.59.381}
Per Bak, Chao Tang, and Kurt Wiesenfeld.
\newblock Self-organized criticality: An explanation of the 1/f noise.
\newblock {\em Phys. Rev. Lett.}, 59(4):381--384, Jul 1987.

\bibitem{benayounploscompbio7e1000486}
Marc Benayoun, Jack~D. Cowan, Wim van Drongelen, and Edward Wallace.
\newblock Avalanches in a stochastic model of spiking neurons.
\newblock {\em PLOS Computational Biology}, 6(7):e1000846--1--e1000846--13,
  2010.

\bibitem{kinouchicopellinaturephys23482006}
Osame Kinouchi and Mauro Copelli.
\newblock Optimal dynamical range of excitable networks at criticality.
\newblock {\em Nature Physics}, 2:348--352, 2006.

\bibitem{shewjon292009}
W.~L. Shew, H.~D. Yang, T.~Petermann, R.~Roy, and D.~Plenz.
\newblock Neuronal avalanches imply maximum dynamic range in cortical networks
  at criticality.
\newblock {\em Journal of Neuroscience}, 29(49):15595--15600, 2009.

\bibitem{newmancp2005}
M.~E.~J. Newman.
\newblock Power laws, pareto distributions and zipf's law.
\newblock {\em Contemporary Physics}, 49(5):323--351, 2005.

\bibitem{clausetsiamr2009}
Aaron Clauset, Cosma~Rohilla Shalizi, and M.~E.~J. Newman.
\newblock Power-law distributions in empirical data.
\newblock {\em SIAM Review}, 51(4):661--703, 2009.

\bibitem{Beggs02062004}
John~M. Beggs and Dietmar Plenz.
\newblock Neuronal avalanches are diverse and precise activity patterns that
  are stable for many hours in cortical slice cultures.
\newblock {\em The Journal of Neuroscience}, 24(22):5216--5229, 2004.

\bibitem{10.1371/journal.pone.0000439}
Alberto Mazzoni, Frédéric~D. Broccard, Elizabeth Garcia-Perez, Paolo Bonifazi,
  Maria~Elisabetta Ruaro, and Vincent Torre.
\newblock On the dynamics of the spontaneous activity in neuronal networks.
\newblock {\em PLoS ONE}, 2(5):e439, 2007.

\end{thebibliography}

\end{document}